\documentclass[12pt]{article}
\usepackage[utf8]{inputenc}

\usepackage{amsmath,amsfonts,graphicx}
\usepackage[square,numbers,sort&compress]{natbib}
\usepackage{fullpage}
\usepackage{lineno}
\usepackage{xcolor}
\usepackage{pdflscape}
\usepackage{hyperref}
\usepackage{multicol,multirow}
\usepackage{authblk}

\newcommand{\edit}{}

\begin{document}

\title{Continuum models describing probabilistic motion of tagged agents in exclusion processes}% Force line breaks with \\

\author[1]{Michael J. Plank}
%\email{michael.plank@canterbury.ac.nz}
\author[2,3]{Matthew J. Simpson}

\affil[1]{School of Mathematics and Statistics, University of Canterbury, Christchurch, New Zealand.}
\affil[2]{School of Mathematical Sciences, Queensland University of Technology (QUT), Brisbane, Australia,}
\affil[3]{ARC Centre of Excellence for the Mathematical Analysis of Cellular Systems, QUT, Brisbane,  Australia.}

\date{}% It is always \today, today,

\maketitle

\begin{abstract}
Lattice-based random walk models are widely used to study populations of migrating cells with motility bias and proliferation. Crowding is typically represented by volume exclusion, where each lattice site can be occupied by at most one agent and conflicting moves are aborted. This framework enables simulations that yield both population-level spatiotemporal agent density profiles and individual agent trajectories, comparable to experimental cell-tracking data. Previous continuum models for tagged-agent trajectories captured trajectory information only, and overlooked any measure of variability.  This is an important limitation since trajectory data is inherently variable.  To address this limitation,  here we derive partial differential equations for the probability density function of tagged-agent trajectories. This continuum description has a clear physical interpretation, agrees well with distributional data from stochastic simulations, reveals the role of stochasticity in different contexts, and generalises to multiple subpopulations of distinct agents.
\end{abstract}

\maketitle
\newpage 

\section{Introduction} \label{sec:Intro}
Lattice-based interacting random walk models, including exclusion processes and generalisations thereof, are often used to study collective motion in populations of biological cells~\cite{Anguige2009,Aubert2006,Bodabilla2019,Callaghan2006,Carrillo2025,Crossley2023,Khain2007,Khain2011,mort2016reconciling,Penington2011,Owen2020,Painter2002,Pillay2017,Pillay2018,Volkening2015}.  Similar lattice-based and lattice-free stochastic models that incorporate crowding mechanisms have been used to study animal and plant dispersal in the context of spatial ecology~\cite{Bolker1997,lewis2000spread,Iwasa2000,Law2003,Cantrell2004}.  Lattice-based exclusion process models enforce each lattice site to be occupied by, at most, a single agent to capture empirically observed crowding effects~\cite{Liggett1999,Plank2025}.  Potential motility or proliferation events that would place more than one agent at the same site are aborted. These models can be used to mimic data generated during cell biology experiments by generating simulation-based snapshots and movies that are directly comparable with experimental images~\cite{Bowden2013}.   These simulation-based outputs are characterised by fluctuations and stochasticity that are also present in experimental data. Beyond simply generating simulation-based images and movies, interacting random walk models can be used to generate ensemble data by considering a suite of identically prepared realisations and averaging over these stochastic realisations to give an averaged, population-level description~\cite{Plank2025}.  This kind of ensemble data can also be modelled using partial differential equation (PDE)-based descriptions for the average population density that are obtained by applying the mean-field approximation~\citep{Baker2010}.  

\begin{figure}
\centering
\includegraphics[width=1.0\linewidth]{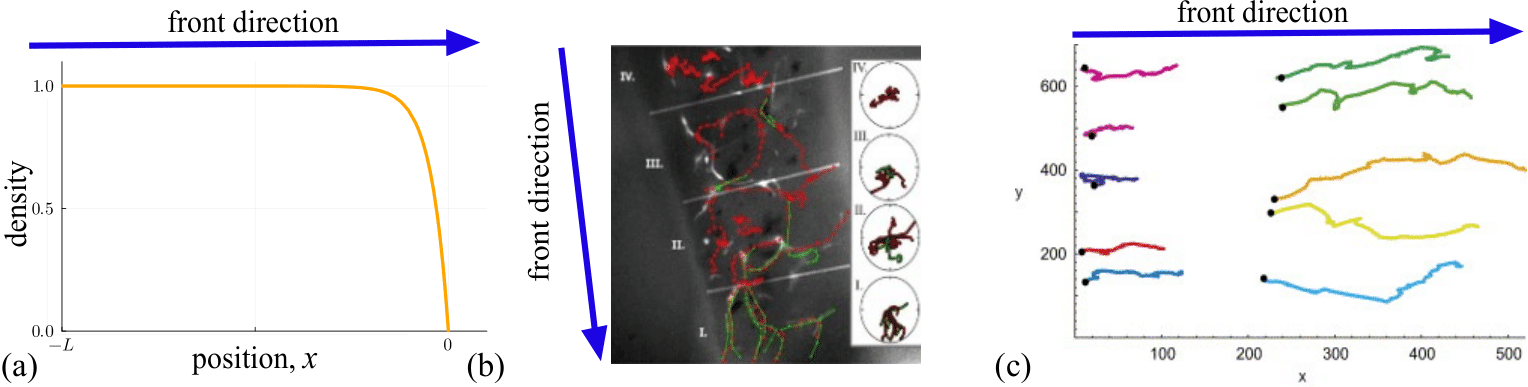}
%\captionsetup{font={small,stretch=0.9}, justification=justified}
\caption{{\edit{\bf Schematic and experimental motivation.} (a) Schematic scaled density profile showing the spatial expansion of a population of cells, undergoing migration and proliferation, leading to the macroscopic propagation of a density front in the positive $x$-direction. (b) \textit{In vivo} tagged cell trajectories reported by Druckenbrod and Epstein~\cite{Druckenbrod2006}.  The direction of the population density front motion is shown with the blue arrow and individual cell trajectories within the population are given by the red and green traces. (c) \textit{In vitro} tagged cell trajectories in a wound healing experiment reported by Cai et al.~\cite{Cai2007}.  The direction of the population density front motion is shown with the blue arrow and individual trajectories show that cells at the edge of population front are biased to move in the same direction as the population front. All images are reproduced with permission.} }
\label{Fig1}
\end{figure}

{\edit Experimental observations often encompass data across multiple scales.   Populations of simulated individuals undergoing motility and carrying capacity-limited proliferation typically lead to moving population fronts, as illustrated by the schematic front in Figure \ref{Fig1}(a) that moves in the positive $x$-direction as the population invades adjacent regions.  Both \textit{in vivo} and \textit{in vitro} experimental observations exhibit this kind of population front movement as illustrated in Figure \ref{Fig1}(b)-(c)~\cite{Druckenbrod2006,Cai2007}.  For example, Figure \ref{Fig1}(b) shows an experimental image of neural crest cells (NCC) cells moving within live, intact tissues during the development of the enteric nervous system (ENS). Individual NCCs are motile and proliferative, leading to a moving front as the population invades adjacent gut tissue~\cite{Druckenbrod2006}.  This experiment involves tagging a small subset of NCCs with fluorescent labels within four different regions of the three-dimensional tissue (Regions I--IV in Figure~\ref{Fig1}(b)).  Region I is at the leading edge of the invading population, whereas Region IV is furthest behind the leading edge.  Superimposing each recorded trajectory within the particular region, shown in the insets labelled I--IV, shows that cells in region I undergo random biased motion where the trajectories are directed in the same direction as the macroscopic wavefront motion.  In contrast, cells in Regions II, III and IV appear to follow random, undirected trajectories.  While it is well understood that NCCs are both motile and proliferative~\cite{Druckenbrod2006}, the precise details of the motility mechanism remain relatively poorly understood.    The experimental image in Figure \ref{Fig1}(c) shows trajectories obtained from a scratch assay, where motion of fibroblast cells is constrained to a two-dimensional plastic substrate~\cite{Cai2007}.  Here, in the simpler two-dimensional geometry, the motion of individual cells can be tracked more easily, giving rise to the trajectories shown in this image.   In this experiment, fibroblast cells undergo combined migration and proliferation, leading to the macroscopic expansion of the population, and the trajectories show that cells towards the leading edge are biased to move in the same direction as the front. While the experimental images and data contained within Figure \ref{Fig1} highlight the importance of labelling and tracking, these data alone does not provide quantitative insight into the underlying mechanisms that drive these experimental observation.  Here we develop quantitative mathematical and computational tools that are motivated by these experimental observations with the aim of developing new modelling tools that can be used to provide insight into these types of experiment.}

In addition to deriving macroscopic, population-level PDE descriptions of the agent density, Simpson et al. (2009)~\cite{Simpson2009b} derived macroscopic models that approximately describe the mean and variance of the locations of tagged agents within the population,  $(x(t),y(t))$. Although this approximately captured variability in agent locations, it did not capture full distributional information as it only tracked the first and second moments of the distribution. Furthermore, the variance approximation ignored temporal correlations in agent locations, which could affect how their distribution evolves over time.   This is an important limitation of the previous approach because trajectory data are known to be highly variable~\cite{Bowden2013}. It is therefore important to develop mathematical tools that can describe both the expected trajectory as well as providing distributional information about the location of tagged agents at a given time.  

In the present study we take a different approach and derive a different, more informative macroscopic PDE that describes the evolution of the probability density function (PDF) for tagged agents $P^{(s)}(x,y,t) \ge 0$, giving a probabilistic prediction of their movement.  The general framework applies to a population of agents composed of $S \in \mathbb{Z}^+$ distinct subpopulations of agents on the lattice, where each subpopulation undergoes biased motility and proliferation at potentially distinct rates.  The system of PDEs for $P^{(s)}(x,y,t) \ge 0$ for $s=1,2,\ldots,S$ can be written as a system of conservation equations with a natural physical interpretation that we will explore. We obtain numerical solutions of the system of PDEs, giving estimates of $P^{(s)}(x,y,t) \ge 0$ for $s=1,2,\ldots,S$, and show that these solutions accurately describe stochastic simulation data, including both the average trends and observed variability in the trajectory data.  {\edit Our study focuses on comparing synthetic data from the stochastic model with solutions of the PDE. While strongly motivated by experimental conditions such as those shown in Figure~\ref{Fig1}, the derivation of a tractable PDE model and validation against synthetic data is an important contribution in itself, providing an analytical framework for applying to experimental trajectory data. }

{\edit Our approach has some parallels with methods, such as Kriging, Brownian bridges and kernel density estimators, used in animal movement ecology to infer trajectories or home-range estimates from telemetry data \cite{fleming2016estimating,fleming2017new}. However, our focus is different in that we specifically account for interactions between multiple individuals in a population, rather than assume they are moving independently, and for proliferation of individuals.   }

\section{Model}\label{sec:Methods}

{\edit In this section, we define a stochastic, agent-based model (ABM) for a population of cells undergoing movement and proliferation in a two-dimensional space. This could represent a specific experimental setup, such as that shown in Figure~\ref{Fig1}b-c. We assume that the cells move on a regular lattice and that crowding effects mean that no more than one cell can occupy a single lattice site. The population may consistent of one or more than subpopulation, where each subpopulation is characterised by four parameters: motility rate, proliferation rate, horizontal direction bias, and vertical directional bias. 

From the ABM, we derive an approximate continuum description that takes the form of a set of reaction-advection-diffusion PDEs for mean agent density. The PDEs are nonlinear due to the interactions between agents. The ABM and corresponding PDEs for agent density have been studied previously \cite{Simpson2009a,Simpson2009b}. The novel contribution of this work is to derive a corresponding PDE for the PDF for the locations of a subset of tagged agents with the population(s). }

\subsection{Stochastic model and continuum-limit description: Single species $S=1$} \label{sec:Single_species_stochastic_model}

Consider a stochastic ABM representing an asymmetric exclusion process, simulated using a random sequential update method~\cite{Chowdhury2005} on a two-dimensional regular lattice with spacing
$\Delta$. In each time step, of duration $\tau$, all agents have the opportunity to move with probability $M$.  A motile agent at $(x,y)$ steps to $(x,y \pm \Delta)$ with probability $(1 \pm
\rho_y)/4$, or to $(x \pm \Delta,y)$ with probability $(1 \pm \rho_x)/4$, where $\left| \rho_x \right| \le 1$ and $\left| \rho_y \right| \le 1$.  Here $\rho_x$ and $\rho_y$ are constant bias parameters that control the degree of motility bias, and we note that setting $\rho_x = \rho_y = 0$ leads to unbiased motility.  In each time step, of duration $\tau$, all agents have the opportunity to proliferate with probability $Q$.  A proliferative agent at $(x,y)$ attempts to place a daughter agent at sites $(x \pm \Delta,y)$ or to $(x,y \pm \Delta)$.  Each of the four possible target sites are chosen with equal probability $1/4$.  Any potential motility or proliferation event that would place an agent on an occupied site is aborted.

The stochastic ABM is related to a continuum partial differential equation (PDE). The continuum description is valid in the constrained limit $\Delta \to 0$, and $\tau \to 0$, where $\Delta^2 / \tau$ is held constant. Denoting the average occupancy of site $(i,j)$, averaged over many realisations, by $\langle C_{i,j} \rangle\in[0,1]$, the spatial and
temporal evolution of the corresponding continuous density $C(x,y,t)$ is governed by~\cite{Plank2025,Codling2008,Simpson2025}
\begin{equation} \label{eq:C_PDE}
\frac{\partial C}{\partial t} + \nabla \cdot \mathcal{\mathbf{J}} = \mathcal{S}, 
\end{equation}
where the $(x,y)$ components of the flux $\mathbf{J} = \left ( \mathcal{J}_{x}, \mathcal{J}_{y} \right)$, and the source term are given by
\begin{align}
\mathcal{J}_{x} &= -D\dfrac{\partial C}{\partial x}+v_x C\left(1 -  C \right), \notag \\
\mathcal{J}_{y} &= -D\dfrac{\partial C}{\partial y}+v_yC\left(1 - C \right), \notag \\
\mathcal{S} &= \lambda C(1-C),
\end{align}
where the drift velocity $\mathbf{v}=(v_x,v_y)$, diffusivity $D$ and proliferation rate $\lambda$ are given by 
\begin{align}
v_x &= \lim_{\Delta ,\tau \to 0}
\left(\frac{M \rho_x \Delta}{2\tau}\right), \, v_y =
\lim_{\Delta ,\tau \to 0} \left(\frac{M \rho_y
 \Delta}{2\tau}\right), \notag \\
D &= \lim_{\Delta ,\tau \to 0}
\left(\frac{M \Delta^2}{4\tau}\right),  \quad \lambda = \lim_{\Delta ,\tau \to 0}
\left(\frac{Q}{\tau}\right). 
\end{align}

\subsection{Stochastic model and continuum-limit description: Multiple species $S>1$} \label{sec:_multiple_species_stochastic_model}
We now generalise the single species model to deal with a total population composed of $S$ subpopulations.  ABM simulations are performed on the same two-dimensional square lattice with spacing
$\Delta$. In each time step, of duration $\tau$, agents from subpopulation $s$ have the opportunity to move with probability $M^{(s)}$.  A motile agent at $(x,y)$ steps to $(x,y \pm \Delta)$ with probability $(1 \pm
\rho_y^{(s)})/4$, or to $(x \pm \Delta,y)$ with probability $(1 \pm \rho_x^{(s)})/4$, where $\left| \rho_x^{(s)} \right| \le 1$ and $\left| \rho_y^{(s)} \right| \le 1$ for each subpopulation, $s=1,2,\ldots,S$.  In each time step of duration $\tau$ agents from the $s$th subpopulation have the opportunity to proliferate with probability $Q^{(s)}$.  In each time step, of duration $\tau$, all agents have the opportunity to proliferate with probability $Q$.  A proliferative agent at $(x,y)$ attempts to place a daughter agent at sites $(x \pm \Delta,y)$ or to $(x,y \pm \Delta)$.  Each of the four possible target sites are chosen with equal probability $1/4$.  Any potential motility or proliferation event that would place an agent on an occupied site is aborted.

The simple exclusion process is related to a continuum model that takes the form of a system of $S$ PDEs in the appropriate limit as $\Delta \to 0$
and $\tau \to 0$~\cite{Simpson2009a}. Denoting the average occupancy of agents from subpopulation $s$ at site $(i,j)$, averaged over many realisations, by $\langle C_{i,j}^{(s)} \rangle\in[0,1]$, the spatial and
temporal evolution of the corresponding continuous density $C^{(s)}(x,y,t)$ is governed by
\begin{equation} \label{eq:C_PDE_multi_species}
\frac{\partial C^{(s)}}{\partial t} + \nabla \cdot \mathcal{\mathbf{J}}^{(s)} = \mathcal{S}^{(s)}, \, \textrm{for} \, s=1,2,\ldots,S
\end{equation}  
where the $(x,y)$ components of the flux $\mathbf{J}^{(s)} = \left ( \mathcal{J}_{x}^{(s)}, \mathcal{J}_{y}^{(s)} \right)$, and the source term for the $s$th subpopulation are given by
\begin{align} \label{eq:C_multispecies_flux}
\mathcal{J}_{x}^{(s)} &= -D^{(s)}\left(1 - T  \right)\dfrac{\partial C^{(s)}}{\partial x}-D^{(s)}C^{(s)}\dfrac{\partial T}{\partial x}+ v_x^{(s)}C^{(s)}\left(1 -  T \right), \notag \\
\mathcal{J}_{y}^{(s)} &= -D^{(s)}\left(1 - T \right)\dfrac{\partial C^{(s)}}{\partial y} -D^{(s)}C^{(s)} \dfrac{\partial T}{\partial y}+ v_y^{(s)}C^{(s)}\left(1 - T \right), \notag \\
\mathcal{S}^{(s)} &= \lambda^{(s)}C^{(s)}(1-T),
\end{align}
for  $s=1,2,\ldots, S$, and where $\displaystyle{T(x,y,t) = \sum_{s=1}^{S} C^{(s)}(x,y,t)}$ is the total density.  Here the drift velocity $\mathbf{v}^{(s)}$, diffusivity $D^{(s)}$ and proliferation rate $\lambda^{(s)}$ for subpopulation $s$ are given by 
\begin{align}
v^{(s)}_x &= \lim_{\Delta ,\tau \to 0}
\left(\frac{M^{(s)} \rho_x^{(s)} \Delta}{2\tau}\right), \, v_y^{(s)} =
\lim_{\Delta ,\tau \to 0} \left(\frac{M^{(s)} \rho_y^{(s)}
 \Delta}{2\tau}\right), \notag \\
D^{(s)} &= \lim_{\Delta ,\tau \to 0}
\left(\frac{M^{(s)} \Delta^2}{4\tau}\right),  \quad \lambda^{(s)} = \lim_{\Delta ,\tau \to 0}
\left(\frac{Q^{(s)}}{\tau}\right). 
\end{align}

{\edit Note that, in general, Eqs. \eqref{eq:C_PDE} and \eqref{eq:C_PDE_multi_species} are nonlinear PDEs as there are nonlinear terms in the expressions for the flux ${\bf J}$ and proliferation rate $\mathcal{S}$. These arise due to the interactions between agents representing crowding effects. In the special case where there is no directional bias and no proliferation (i.e. $v_x=v_y=\lambda=0$), Eq. \eqref{eq:C_PDE} for the single-species cases reduces to the linear diffusion equation, but Eq. \eqref{eq:C_PDE_multi_species} for the multi-species cases is still nonlinear \cite{Simpson2009a}.}

\subsection{Tagged agents}
We now describe how to obtain a macroscopic model describing the motion of individual agents within the population using a novel probabilistic framework.  To proceed, we suppose that one of the individual agents in the population is {\em tagged} at time $t=0$ and its location tracked through time. Let $P(x,y,t)$ denote the PDF for the agent's location at time $t$. We now derive a continuum-limit PDE for $P(x,y,t)$ via a similar method to that used to derive the PDE for $C(x,y,t)$. 

To begin, let $P_{i,j}(t)$ denote the probability that the tagged agent is located at lattice site $(i,j)$ at time $t$. As previously, let $C_{i,j}(t)$ denote the probability that the $(i,j)$ lattice site is occupied by any agent at time $t$ so that $C_{i,j}(t)=0$ indicates that site $(i,j)$ is vacant and $C_{i,j}(t)=1$ indicates that site $(i,j)$ is occupied.  With this framework we may write a discrete conservation equation for $P_{i,j}(t)$ as follows,
\begin{align} \label{eq:cons_statement}
 &P_{i,j}(t+\tau) = \overbrace{P_{i,j}(t)\left(1- M \right)}^{\textrm{agent is at } (i,j) \textrm{ and does not attempt to move}}  \\
&+\underbrace{\dfrac{M P_{i,j}(t)}{4}\left[ (1+\rho_x) C_{i+1,j}(t) + (1-\rho_x) C_{i-1,j}(t) + (1+\rho_y)C_{i,j+1}(t) + (1-\rho_y)C_{i,j-1}(t)\right]}_{\textrm{agent is at } (i,j) \textrm{ and attempts to move but is unsuccessful due to crowding}} \notag \\
&+\underbrace{\dfrac{M \left[1-C_{i,j}(t)\right]}{4}\left[ (1+\rho_x) P_{i-1,j}(t) + (1-\rho_x)P_{i+1,j}(t) + (1+\rho_y)P_{i,j-1}(t) + (1-\rho_y)P_{i,j+1}(t) \right].}_{\textrm{agent is at } (i \pm 1,j) \textrm{ or } (i,j \pm 1) \textrm{ and successfully moves to } (i,j) }  \notag 
\end{align}
Like the derivation of the PDE for $C(x,y,t)$, this conservation statement invokes a standard mean-field approximation, which assumes that the occupancy status of adjacent lattice sites are independent random variables \cite{Baker2010}. 

 The next step in obtaining the continuum-limit description is to identify discrete quantities $P_{i,j}(t)$ and $C_{i,j}(t)$ with smooth functions $P(x,y,t)$ and $C(x,y,t)$, respectively.  To proceed, we replace terms of the form $C_{i\pm 1,j}(t)$,  $C_{i,j\pm 1}(t)$, $P_{i\pm 1,j}(t)$ and  $P_{i,j\pm 1}(t)$ in Eq. \eqref{eq:cons_statement} with standard two-dimensional Taylor expansions about $(x,y)$.  
Taking the limit as $\Delta \to 0$ and $\tau \to 0$ with the ratio $\Delta^2/\tau$ held constant~\cite{Codling2008,Plank2025}, terms of order $\mathcal{O}(\Delta^3)$ vanish leading to the following conservation PDE  
\begin{equation} \label{eq:P_PDE}
\frac{\partial P}{\partial t} +\nabla \cdot \mathcal{\mathbf{J}}=0,
\end{equation}
where the components of the flux $\mathbf{J} = \left ( \mathcal{J}_{x}, \mathcal{J}_{y} \right)$ are given by
\begin{align}
\mathcal{J}_{x} &= -D\left(1 -C\right)\dfrac{\partial P}{\partial x}-DP\dfrac{\partial C}{\partial x}+ v_xP\left(1 -  C \right), \notag \\
\mathcal{J}_{y} &= -D\left(1 - C \right)\dfrac{\partial P}{\partial y} -DP\dfrac{\partial C}{\partial y}+ v_yP \left(1 - C \right). 
\end{align}
These three terms in these expressions for the components of flux have intuitive mechanistic interpretations. The first term is a {\em self-diffusion} term for $P$ that is attenuated by a factor of $(1-C)$ due to crowding effects.  The second term is an advection-like flux term representing the transport of $P$ in proportion to the macroscopic diffusive flux $-D\nabla C$, sometimes called {\em collective diffusion}.  The third term is an advective flux represents transport of $P$ with advection velocity $v_{x}$ attenuated by a factor of $(1-C)$ due to crowding.  In the low-density limit where $C \to 0^+$ and $\nabla C = (0,0)$, these expressions for the flux simplify to a standard advection-diffusion flux, giving $\mathcal{J}_{x} = -D \partial P / \partial x + v_xP$ and $\mathcal{J}_{y} = -D \partial P / \partial y + v_yP$~\cite{Codling2008}. 

The derivation of the multi-species case follows the same logic, leading to  
\begin{equation} \label{eq:P_PDE_multi_species}
\frac{\partial P^{(s)}}{\partial t} +\nabla \cdot \mathcal{\mathbf{J}}^{(s)}=0, 
\end{equation}
for subpopulation $s=1,2,\ldots, S$, and the components of the flux $\mathbf{J}^{(s)} = \left ( \mathcal{J}_{x}^{(s)}, \mathcal{J}_{y}^{(s)} \right)$ are given by
\begin{align} \label{eq:P_multispecies_flux}
\mathcal{J}_{x}^{(s)} &= -D^{(s)}\left(1 -T\right)\dfrac{\partial P^{(s)}}{\partial x}-D^{(s)}P^{(s)}\dfrac{\partial T}{\partial x}+ v_x^{(s)} P^{(s)}\left(1 -  T \right), \notag \\
\mathcal{J}_{y}^{(s)} &= -D^{(s)}\left(1 -T\right)\dfrac{\partial P^{(s)}}{\partial y} -D^{(s)}P^{(s)}\dfrac{\partial T}{\partial y}+ v_y^{(s)} P^{(s)} \left(1 - T \right), 
\end{align}
where $\displaystyle{T(x,y,t) = \sum_{s=1}^{S} C^{(s)}(x,y,t)}$ is the total density.  The three terms in these expressions for the multi-species flux have a similar physical interpretation as for the single species case described above, and when $S=1$ the multi-species PDE model reduces to the single species case given by Eq. \eqref{eq:P_PDE}. {\edit Like the PDEs for agent density $C$, Eqs. \eqref{eq:P_PDE} and \eqref{eq:P_PDE_multi_species} are both nonlinear PDEs due to the nonlinearities in the flux ${\bf J}$.}

It is worth noting the parallels between the flux terms for the PDF for tagged agent location in Eq. \eqref{eq:P_multispecies_flux} and those for macroscopic agent density in Eq. \eqref{eq:C_multispecies_flux}. Effectively, a tagged agent can be interpreted as a separate species, moving within a population with total macroscopic density $T(x,y,t)$. However, there are important differences between the variables $C^{(s)}$ and $P^{(s)}$.  Since we have an exclusion process, a key property of the ABM is that the occupancy of individual sites cannot exceed $1$, and in the continuum limit this means that we have $C^{(s)}(x,y,t) \in [0,1]$ for $s=1,2,\ldots,S$, with the additional constraint that  $\displaystyle{T(x,t) = \sum_{s=1}^{S} C^{(s)}(x,y,t)}\in [0,1]$. These properties do not hold for the PDFs for tagged agent locations since we have $P^{(s)}(x,y,t) \ge 0$ with the constraint that $\displaystyle{\int_\Omega P^{(s)}(x,y,t) \, \textrm{d}x \,\textrm{d}y} = 1$, for $s=1,2,\ldots,S$. There is no upper bound on $P^{(s)}$ and indeed to model a tagged agent initially located at $(x,y)=(x_0,y_0)$, we use an initial condition $P^{(s)}(x,y,0) = \delta(x_0,y_0)$ where $\delta(\cdot)$ is the Dirac delta function.

{\edit For details of the methods used to simulate the ABM and numerically solve the PDE model, see Supplementary Material \cite{supplementary_material}.}
Matlab software to reproduce the results in this study is publicly available at~\cite{plank_tagged_agents_repo}.  We encourage readers can use this software directly to replicate the results presented in this study, or to adapt the software and explore different scenarios, such as working with different initial conditions, parameter values, simulation durations.

\section{Results} \label{sec:results}
{\edit In section \ref{sec:Methods}, we presented the model derivation for the general multi-species case ($S>1$) in two spatial dimensions in order to provide a comprehensive theory. In this section, we provide numerical comparisons of stochastic ABM simulations and solutions of the PDE model, focusing on the single-species case ($S=1$) and the situation where lattice occupancy is independent of the $y$ coordinate (see Supplementary Material \cite{supplementary_material} for details). We therefore drop the superscript $(s)$ and use $C(x,t)$ and $P(x,t)$ respectively to denote average agent density and PDF for tagged agents at location $x$ and time $t$.} We systematically explore the behaviour of the models with and without direction bias, and with and without agent proliferation. {\edit Our results illustrate the qualitative types of behaviour that can occur in different situations and test the accuracy of the PDE model approximation $P(x,t)$ for the distribution of tagged agent locations in stochastic simulations.} 

\subsection{Unbiased motility and no proliferation}
When there is no bias or proliferation, the macroscopic agent density $C(x,t)$ evolves over time according to the linear diffusion equation (Figure \ref{fig:all_cases}a) and we see that averaged simulation data matches the solution of the mean-field PDE very well. Tagged agents initially near either leading edge of the population develop a skewed distribution in the direction away from centre of the the initial population distribution at $x=0$ (Figure \ref{fig:all_cases}b, blue and yellow). This skewing of the distribution is caused by crowding effects, meaning that motility events towards $x=0$ are more likely to be aborted than motility events away from $x=0$.  The net result of these crowding effects is that the tagged agents tend to drift down the macroscopic density gradient. Tagged agents initially in the centre of the population $x=0$ remain symmetrically distributed about $x=0$ since there is no macroscopic density gradient here owing to symmetry (Figure \ref{fig:all_cases}b, red). These tagged agents diffuse more slowly due to the higher local density, meaning that these agents experience a higher probability of aborted moves than tagged agents at the leading edges of the population.  In all cases, the solution of the PDE model gives a good approximation to the observed distribution of tagged agent location (including its asymmetric shape) at the end of the simulation (Figure \ref{fig:all_cases}b), as well as capturing the dynamics of the observed distribution over time (Figure \ref{fig:all_cases}c).

\subsection{Biased motility and no proliferation}

\begin{figure}
    \centering
    \includegraphics[trim={2.3cm 1.9cm 2.2cm 1cm},clip,width=1.0\linewidth]{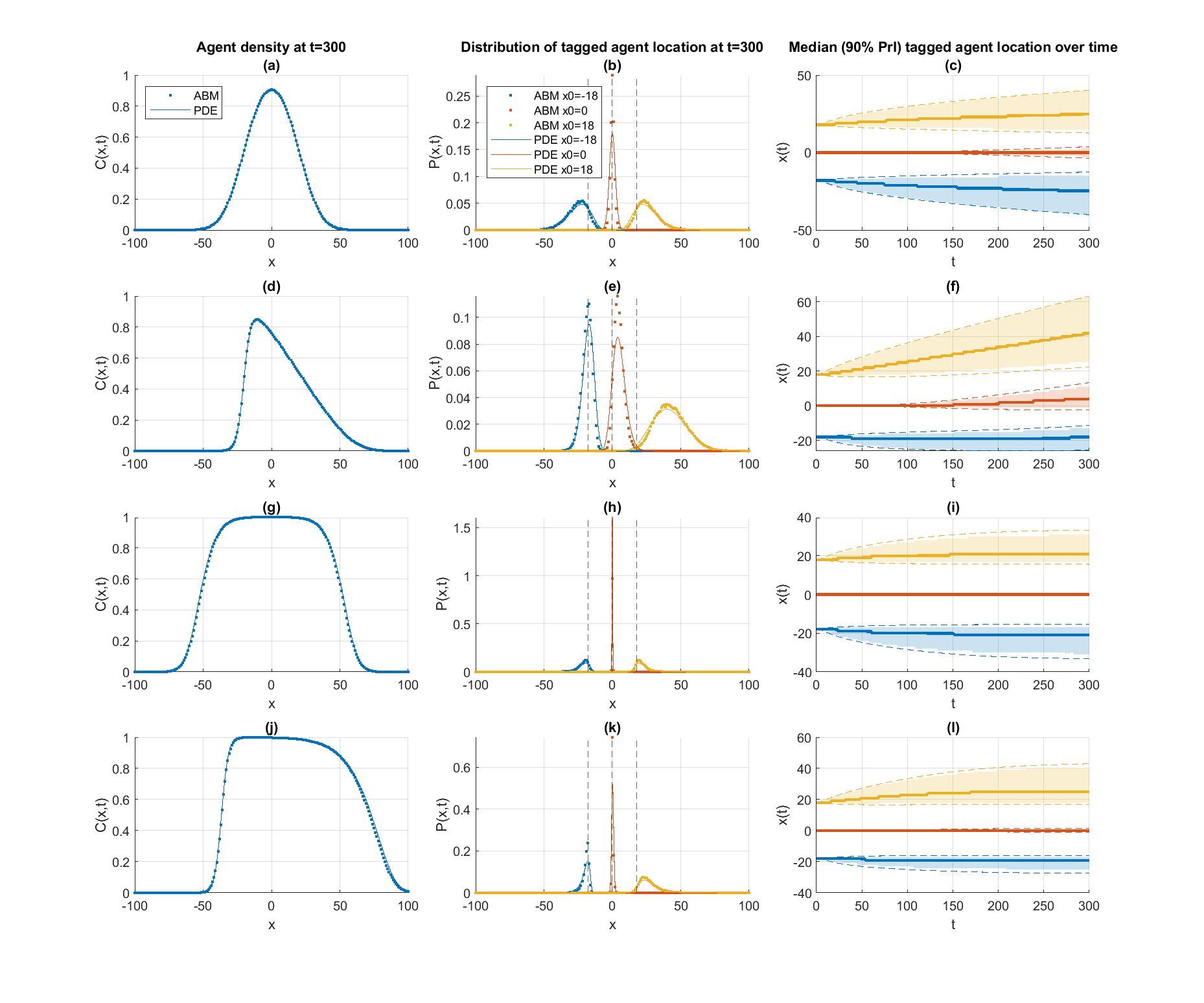}
%      \captionsetup{font={small,stretch=0.9}, justification=justified}
    \caption{{\bf Comparison of discrete and continuum models for a motile population:} (a)-(c) without bias or proliferation; (d)-(f) with bias and without proliferation; (g)-(i) without bias and with proliferation; (j)-(l) with bias and proliferation. Left column of plots show agent density $C(x,t)$ at $t=300$; middle column shows distribution of the location at $t=300$ of tagged agents initially located near the left-hand leading edge ($x_0=-18$, blue), in the centre of the population ($x_0=0$, red) and near the right-hand leading edge ($x_0=18$, yellow); right column shows the median and 90\% PrI of tagged agent locations over time time according to the ABM (thick solid curve = median, shaded band = 90\% PrI) and the PDE (thin solid curve = median, dashed curves = 90\% PrI). Vertical dashed lines in middle column show the initial location of tagged agents. Discrete parameter values $M = 1$ and $\rho_x=Q=0$ corresponding to  $D=0.25$ and $v=\lambda=0$ for a simulation with $\Delta = \tau = 1$.}
    \label{fig:all_cases}
\end{figure}

When there is a directional bias in movement in the positive $x$ direction with $\rho_x > 0$, the macroscopic agent density becomes skewed to the right (Figure \ref{fig:all_cases}d), as is well known~\cite{Simpson2009b}.   Again, we see that averaged simulation data for the density matches the solution of the mean-field density, including the skew in the density profile.  The distribution of tagged agents initially near the right-most leading edge drifts in the positive $x$-direction due to the motility bias, and diffuses over time (Figure \ref{fig:all_cases}e, yellow). Tagged agents initially in the centre $x=0$ and at the left-most leading edge drift more slowly and do not spread out as rapidly due to crowding effects (Figure \ref{fig:all_cases}e, red, blue). Again, the PDE gives a good approximation to the distribution of tagged agent locations in the ABM (Figure \ref{fig:all_cases}e-f).

\subsection{Unbiased motility and proliferation}

Including proliferation in the ABM, with no directional bias, means that the total population size increases over time and the mean-field PDE model is the well-known Fisher-Kolmogorov model~\cite{EdelsteinKeshet2005,Kot2003,Murray2002}.   Therefore, after a sufficient duration of time the macroscopic agent density eventually approaches a travelling wave profile with $C(x,t) \to 1^-$ (representing a fully occupied lattice) behind the wave front and $C(x,t) \to 0^+$ ahead of the leading edge (Figure \ref{fig:all_cases}g). The distributions of tagged agents initially close to either leading edge start to spread out at the beginning of the simulation, and eventually become skewed away from the centre (Figure \ref{fig:all_cases}h, blue and yellow), as occurred in the case without proliferation in Figure \ref{fig:all_cases}a-c. However, these agents that are initially at the leading-ledge of the population eventually become relatively immobile and fixed in place once the lattice in their local neighbourhood becomes fully occupied. As a result, there is little further change in their distributions after approximately $t=150$ (Figure \ref{fig:all_cases}i). The tagged agents initially located in the centre of the population near $x=0$ are unable to move at all and so retain their initial Dirac delta-like distribution for the whole simulation (Figure \ref{fig:all_cases}h-i, red). Again, the solution of the PDE model for $P(x,t)$ captures these trends and dynamics reasonably accurately.

\subsection{Biased motility and proliferation}

In simulations with both directional bias and proliferation, the macroscopic agent density develops an asymmetric profile, with the travelling wave at the right-most leading edge moving faster than that at the left-most leading edge (Figure \ref{fig:all_cases}j). Tagged agent distributions display varying degrees of drift and diffusion depending on their initial location within the population (Figure \ref{fig:all_cases}k). Tagged agents in the centre are unable to move and are fixed in place almost immediately due to high local density (Figure \ref{fig:all_cases}k-l, red).  Those tagged agents initially near the left-hand edge spread out until around $t=150$ when they become immobile (Figure \ref{fig:all_cases}k-l, blue), whereas the tagged agents initially near the right-most leading edge are able to move more rapidly and their distribution is continuing to spread out at the end of the simulation at $t=300$ (Figure \ref{fig:all_cases}k-l, yellow). This is because some of these tagged agents are able to remain ahead of the wave front in a region where macroscopic density remains sufficiently small. Again the PDE provides a good approximation to the ABM results.

\subsection{Accuracy of the approximation for tagged agents}

We explored how the accuracy of the PDE model compared to the method proposed by Simpson, Landman and Hughes~\citep{Simpson2009b} for calculating the mean and variance of tagged agent locations. This method, which we refer to as the SLH approximation, relies on calculating the rate of change of the mean and variance for an agent that is positioned at the mean location, and thus ignores the effects of uncertainty in the agent location. In all cases investigated, the PDE model and the SLH approximation both predict the mean agent location very accurately (Supplementary Figures S1a,c,e,g \cite{supplementary_material}). In cases without proliferation, the PDE model predicts the variance as well or better than the SLH approximation (Supplementary Figure S1b,d \cite{supplementary_material}). When there is proliferation, the results are more mixed. When there is no bias, the SLH predicts the variance of agents initially near a leading edge more accurately than the PDE model (Supplementary Figure S1f \cite{supplementary_material}), but the reverse is true when there is bias (Supplementary Figure S1h \cite{supplementary_material}). The PDE model tends to overestimate the variance in tagged locations in all cases, while the SLH approximation underestimates it in some cases and overestimates in others. 

Whilst we have calculated the SLH approximation for all cases investigated here for completeness, it should be noted that Simpson et al.~\citep{Simpson2009b} did not consider proliferation, and only examined agents initially located near the right-most leading edge of the population.  Also the SLH approximation provides no information about tagged agent locations beyond their mean and variance, whereas our model provides access to the full distribution via its PDF. {\edit This provides a richer set of information, which enables predictions and analyses that would not otherwise be possible. For example, our results could be used to define a tractable likelihood function for observed data on tagged agent locations. This would enable parameter inference and uncertainty quantification via either frequentist or Bayesian methods \cite{Simpson2025}.   }

In circumstances where the PDE model for $P$ loses accuracy, this is likely due to failure of the mean-field approximation to hold, meaning there are non-negligible correlations in the occupancy status of neighbouring lattice sites~\cite{Baker2010}. The results we have presented demonstrate that this affects the accuracy in terms of the tagged agent locations, $P$, more than those for the macroscopic density, $C$. This can be explained by the fact that, when two agents are adjacent, the potential for aborted movements by either agent reduces the expected flux equally in both directions, and so the net effect of correlations on the macroscopic density profile is zero. This symmetry does not hold for the tagged agent distribution because it is tracking the location of an individual agent, and so it makes sense that correlations will potentially have a bigger effect. 

To illustrate this, we explored a test case in which all lattice sites were initially occupied with probability $0.5$ and there is no proliferation or motility bias. This means that the macroscopic agent density is uniformly constant at $C(x,t)=0.5$ and the PDE for $P(x,t)$ in Equation \eqref{eq:P_PDE} reduces to the linear diffusion equation with diffusivity $0.5D$. This PDE for $P(x,t)$ would accurately describe a situation where 50\% of the tagged agent's attempted movement events are aborted. However, simulations of the ABM show that the tagged agent's distribution spreads out more slowly than this PDE predicts (Supplementary Figure S2 \cite{supplementary_material}), indicating that more than 50\% of moves are aborted. This is due to the positive correlation between neighbouring lattice sites, which means that, conditional on the tagged agent being at site $(i,j)$, the occupancy probabilities of sites $(i\pm 1,j)$ and $(i,j\pm 1)$ are slightly greater than $0.5$.

\section{Conclusion and Future Work} \label{sec:Conclusion}
In this work, we have investigated lattice-based random walk models of individual-level motility and proliferation mechanisms in the context of an exclusion process framework, where each lattice site can be occupied by no more than a single agent~\cite{Liggett1999}. Stochastic models with exclusion are often used to represent cell biology experiments where crowding effects can be very pronounced~\cite{Anguige2009,Aubert2006,Bodabilla2019,Callaghan2006,Khain2007,Khain2011,mort2016reconciling,Penington2011,Owen2020,Pillay2017,Pillay2018,Volkening2015}. Motivated by experimental observations in Figure \ref{Fig1}, we used a stochastic ABM to describe the spatial evolution of the population-level density profile, as well as considering the motion of a small number of tagged agents within the broader population.  We explored the existing continuum-limit description of the macroscopic density $C(x,y,t) \in [0,1]$ alongside a new continuum-limit description of the PDF for the location of tagged agents $P(x,y,t) \ge 0$.  The derivation of the continuum limit was extended to the situation where we have $S \in \mathbb{N}^+$ distinct subpopulations on the lattice so that the dependent variables in the PDE models are $C^{(s)}(x,y,t) \in [0,1]$ and $P^{(s)}(x,y,t) \ge 0$ for $s=1,2,3,\ldots,S$.  Repeated stochastic simulation data showed that the continuum-limit model for $C$ and $P$ provide a good match data obtained from the computationally expensive stochastic simulations.  In particular, we showed that numerical estimates of $P$ provide a probabilistic interpretation of the motion of tagged agents since the continuum-limit model can be used to predict both the expected location of the tagged agents, as well as predicting the variability in their location.  Numerical tests confirmed that the solution of the continuum-limit model for $P$ provides a reasonable match to data from the stochastic ABM under a range of conditions including unbiased and biased motility, both in the presence and absence of agent proliferation.   

We have considered particular applications of the discrete and continuum models that focus on the canonical problem of dealing with a single population of agents, $S=1$.  This is consistent with the experimental images in Figure \ref{Fig1} that involve a single population of cells in which both the population-level expansion and the motion of individual tagged cells are measured and reported.  The discrete-continuum comparisons in this work focus the most fundamental scenario where macroscopic gradients vanish in the $y$--direction, and the initial condition for the tagged agent PDE is $P(x,0) = \delta(x_0)$.  This initial condition corresponds to the tagged agent(s) located precisely at $x=x_0$ at $t=0$.  We made this choice since it is arguably the simplest and most natural way to explore discrete simulations to compare with the corresponding solution of the continuum model and it directly mimics the experimental scenario in Figure \ref{Fig1}(c).  This approach, however, makes the strong assumption that the initial location of tagged agents is known precisely, which is potentially untrue in practice.  Alternatively, we may assume that the initial location of tagged agents are contained within some interval by setting $P(x,0) = 1/\ell$ for $x \in [x_1, x_1 + \ell]$ and $P(x,0) = 0$ otherwise, for some location $x_1$ and some interval length $\ell > 0$.  This alternative initial condition assumes that tagged agents are equally likely to be at any location within the interval $x \in [x_1, x_1 + \ell]$.  This approach allows us to introduce some uncertainty into the initial location of the tagged agent(s), which may be more appropriate when modelling real experimental data, and we note that other initial conditions that further generalise these ideas are also possible.

In addition to making various continuum-discrete comparisons for a range of problems involving a single population of agents with $S=1$, we also derived continuum models for the growth and spatial spreading of populations that are composed of more than one distinct subpopulations, $S > 1$.  While we have not made continuum-discrete comparisons for $S > 1$, the tools provided in this work lay the foundation for future comparisons to be made in these cases. We have focused on mathematical models in which individuals undergo motility and proliferation events, which leads to populations of agents that either maintain their size (if $Q=\lambda =0$) or grow over time (if $Q >0, \lambda > 0$).  We chose to focus on these conditions because cell death is often absent from many {\em in vitro} experiments.  It is also possible, however, to extend the discrete models to incorporate different forms of agent death, and the same framework can be used to derive continuum models for $P$, except that care must be taken to deal with the possibility that tagged agents can die, and indeed there is the possibility that entire subpopulations will go extinct during the simulations in these cases.  We leave these extensions of our current modelling framework for future consideration.

The PDE models we derived for the density $C^{(s)}$ involve a source term whenever $Q^{(s)} > 0$ for $s=1,2,\ldots,S$.  The corresponding PDE for the PDF for tagged agents $P^{(s)}$ does not involve any source term, even when the simulation involves agent proliferation with $Q^{(s)} > 0$.  This property is a reflection of the fact that our model tracks the location of agents that are tagged at the beginning of the experiment, without consideration of any offspring they may produce. Our approach could be generalised to track the original tagged agents, as well as any daughter agents and associated lineages they give rise to.  This would lead to a PDE model for $P^{(s)}$ that involves a source term to account for the fact that tagged agents themselves can be involved in proliferation events, similar to mathematical models of lineage tracing~\cite{Cheeseman2014}.

{\edit A potential use case for our framework is particle identification. For example, suppose there are multiple tagged agents that are are observed at snapshots in time, but are not {\em a priori} distinguishable. Our model provides the probability that an agent which was initially at known location ${\bf x}_0$ will be located at location ${\bf x}$ at time $t$. This could be used to provide a probabilistic estimate of an observed agent's identity, but we leave this question for future work.  }
Another approach for generalising our mathematical models is to recast the stochastic lattice-based random walk model into a stochastic lattice-free random walk~\cite{Bruna2012,Bruna2012JCP,Plank2012}.  The main advantage of working in a lattice-free framework is that the motility direction is continuous rather than being discrete, however this additional flexibility comes at the cost of additional mathematical complexity, which makes deriving appropriate mean-field PDE descriptions more difficult~\cite{Bruna2012,Bruna2012JCP,Plank2012}. 

 Regardless of whether we work with a lattice-based or lattice-free framework, deriving appropriate continuum limits for the population-level density and individual-level tagged agent trajectory properties provides an opportunity to perform parameter inference using simultaneous observations of the agent density, $C^{(s)}$, and the location of tagged agents, $P^{(s)}$ for $s=1,2,\ldots,S$.  Having simple computational tools that enable us to understand how $C^{(s)}$ and $P^{(s)}$ vary with the model parameters provides a key ingredient for either Bayesian or frequentist statistical inference~\cite{Hines2014,Wasserman2004,Simpson2026} that makes full use of all available (combined) data. Having a continuum-limit PDE description is highly advantageous for parameter inference as: (1) when coupled with a suitable observation noise model~\citep{Simpson2025,Hines2014}, it provides access to a likelihood function that can be used for optimisation or sampling-based methods; and (2) it is more computationally efficient than generating repeated stochastic ABM simulations, which can be prohibitively expensive for performing parameter inference, especially if the parameter space is high-dimensional.

 {\edit The approach we have taken in this study takes the intrinsic movement, proliferation and interaction rules that define the ABM as given and uses these to make predictions about how population density and agent locations evolve over time. In some situations, these rules may be approximately known from biophysical principles or prior data. In other situations, they may be unknown, in which case models such as those presented here can be used to statistically test alternative hypotheses against observed data, using parameter inference and model selection approaches \cite{Hines2014,Simpson2025,gelman2020bayesian}. Hence, our methods provide an avenue to inferring mechanistic behavioural rules from empirical cell trajectory data.}

\nocite{Donea2003}

\noindent 
\textit{Acknowledgements} This work is partly supported by the Australian Research Council (DP230100025, CE230100001) and the Marsden Fund (24-UOC-020). The authors are grateful to two anonymous reviewers for comments on an earlier version of this article. 

%\bibliography{references}% Produces the bibliography via BibTeX.
\providecommand{\noopsort}[1]{}\providecommand{\singleletter}[1]{#1}%

\clearpage

\renewcommand\theequation{S\arabic{equation}}
\renewcommand\thefigure{S\arabic{figure}}
\renewcommand\thetable{S\arabic{table}}
\renewcommand\thesection{S\arabic{section}}
\renewcommand{\refname}{Supplementary references}
\setcounter{section}{0}
\setcounter{equation}{0}
\setcounter{figure}{0}

\section*{Supplementary Material}

\section{Numerical methods} \label{sec:numerical_methods}
In all cases we consider ABM simulations on a regular lattice with lattice spacing $\Delta = 1$, lattice width $W$ and height $H$, where initially the occupancy status of each lattice site is independent of vertical position~[22].  This approach, together with implementing either periodic or reflecting boundary conditions and assuming there is no bias in the $y$ direction ($\rho_y=0$) reflects the geometry of various experimental measurements illustrated in Figure 1. It also simplifies the PDE models since macroscopic gradients in the $y$ direction vanish, meaning that average density $C^{(s)}(x,y,t)$ is independent of $y$ and $\nabla \cdot \mathbf{J}^{(s)}$ simplifies to $\partial \mathcal{J}_x^{(s)} /\partial x$~[22].  Under these conditions, while individual agents in the discrete simulations are free to move in both the $x$ and $y$-directions, the macroscopic PDE models simplify to PDEs with one spatial dimension.  The PDE for average agent density becomes
\begin{equation}
\frac{\partial C^{(s)}}{\partial t} + \dfrac{\partial \mathcal{J}_x^{(s)}}{\partial x}= \mathcal{S}^{(s)},
\end{equation}
 where
\begin{align}
\mathcal{J}_{x}^{(s)} &= -D^{(s)}\left(1 - T  \right)\dfrac{\partial C^{(s)}}{\partial x}-D^{(s)}C^{(s)}\dfrac{\partial T}{\partial x}+ v_x^{(s)}C^{(s)}\left(1 -  T \right), \notag \\
\mathcal{S}^{(s)} &= \lambda^{(s)}C^{(s)}(1-T),
\end{align}
for  $s=1,2,\ldots, S$, and $\displaystyle{T(x,t) = \sum_{s=1}^{S} C^{(s)}(x,t)}$ is the total density.

A similar simplification holds for the PDE model for the tagged agents, giving
\begin{equation}
\frac{\partial P^{(s)}}{\partial t} +\dfrac{\partial \mathcal{J}^{(s)}}{\partial x}=0, 
\end{equation}
with
\begin{align}
\mathcal{J}_{x}^{(s)} &= -D^{(s)}\left(1 -T\right)\dfrac{\partial P^{(s)}}{\partial x}-D^{(s)}P^{(s)}\dfrac{\partial T}{\partial x}+ v_x^{(s)} P^{(s)}\left(1 -  T \right). 
\end{align}

For all simulations and continuum--discrete comparisons presented in this work, we will focus on capturing properties of the key experimental results in Figure 1 that involve a single population of cells.  Accordingly we will set $S=1$ and drop the superscript on the $C$ and $P$ variables for the presentation and discussion of several examples.  We note, however, that our approach and general trends in our results also hold when these concepts are applied to dealing with multiple subpopulations of agents with $S > 1$.

The mathematical models for $C(x,t)$ and $P(x,t)$ reduce to the following coupled PDE system
\begin{align}
\frac{\partial C}{\partial t} &= D\frac{\partial^2 C}{\partial x^2} - v_x \frac{\partial}{\partial x}\left[ C(1-C) \right], \label{eq:PDE1} \\ 
\frac{\partial P}{\partial t} &= D\left[ (1-C)\frac{\partial^2 P}{\partial x^2} + P \frac{\partial^2 C}{\partial x^2} \right] - v_x \frac{\partial}{\partial x}\left[ P(1-C) \right], \label{eq:PDE2} 
\end{align}
with no-flux boundary conditions for $C(x,t)$ and $P(x,t)$, at $x=\pm W/2$.   We solve this system of coupled PDEs using the method of lines by uniformly discretising the $x$ coordinate with a constant step size $h$ and approximating the spatial derivatives in the PDE models using finite differences. For the second-order diffusive derivatives in Equations \eqref{eq:PDE1}--\eqref{eq:PDE2}, we use the standard central difference approximation. For the first-order advective derivatives, we use an upwind scheme, which approximates the derivative of a function at $x$ in terms of the values of the function at $x$ and at $x-h$ for $v_x > 0$ or values of the function at $x$ and $x+h$ for $v_x < 0$.  We consider the initial condition 
\begin{align}\label{eq:ic} 
C(x,0) &= \left\{ \begin{array}{ll}  C_0, & |x| \le  a,  \\ 0, & \textrm{otherwise}, \end{array} \right. \\
P(x,0) &= \left\{ \begin{array}{ll} \dfrac{1}{h}, & x=x_0, \\ 0, & \textrm{otherwise}, \end{array} \right.
\end{align}
where the initial condition for $P(x,0)$ is a standard numerical approximation to the Dirac delta function~[47]. The resulting system of ordinary differential equations was solved using the built-in Matlab solver {\em ode45}.

To visualise the width of the distribution in both the stochastic ABM and the PDE, we calculate the $q\mathrm{th}$ quantile of the tagged agent distribution $P(x,t)$ at time $t$, i.e. the value of $x_q$ satisfying $F(x_q,t)=q$, where $F(x,t)$ is the cumulative distribution function of $P(x,t)$ defined by
\begin{equation}
   F(x,t)= \int_{-W/2}^{x} P(x',t) \, \textrm{d}x'. 
\end{equation}
In practice, we compute $x_q$ numerically by firstly approximating the integral for $F(kh,t)$ as the discrete sum $\displaystyle{ h\sum_{i=1}^k P(x_i,t)}$ for $k=1,2,\ldots,W/h$, and then using linear interpolation to evaluate $F(x,t)$. Here $h$ is the same mesh spacing that we use to approximate the spatial derivatives in the finite difference approximation.  Note that the width $W$ of the computational domain was taken to be sufficiently large that the value of $P(x,t)$ is numerically very close to zero at both boundaries, $x=\pm W/2$, for all problems considered. Therefore, different methods for numerical quadrature, such as the trapezium rule, would gives result almost identical to the approximation we used.  In this work we characterise the location of tagged agents by computing the median of the distribution by setting $q=0.5$, and we characterise the width of the distribution by calculating the 90\% probability interval (90\% PrI) by setting $q=0.05$ and $q=0.95$.  In summary, this approach allows us to take numerical solutions for $P(x,t)$ and compute both a median position and a probability interval, which gives us a simple way of reporting trajectory data that is consistent with experimental data and with stochastic simulation data.

All simulations use a lattice of width $W=300$, height $H=100$, and initial density $C_0=1$ in the region $|x|\le 20$, i.e. Equation \eqref{eq:ic} with  $a=20$.   In each ABM simulation, we place a set of 10 tagged agents at each of three initial locations: $x_0=-18$ (close to the left-most leading edge); $x_0=0$ (in the centre of the population); and $x_0=18$ (close to the right-most leading edge). The initial $y$-coordinate of the tagged agents is chosen randomly (without replacement) from $y=0,1,\ldots, H$.   This design of the placement of agents replicates the experimental design taken by Cai et al.~[26] shown in Figure 1(c) where five cells are chosen along each vertical transect with constant horizontal position, $x$.   To generate averaged data we perform $5000$ independent realisations of the stochastic ABM, and calculate the average column occupancies, and the proportion of each group of tagged agents in each column, over all identically prepared realisations.

Matlab software to reproduce the results in this study is publicly available at~[35]. All analyses were run in Matlab R2022b.

\newpage 

\begin{figure}
    \centering
    \includegraphics[trim={2.3cm 1.5cm 0.5cm 1cm},clip,width=\linewidth]{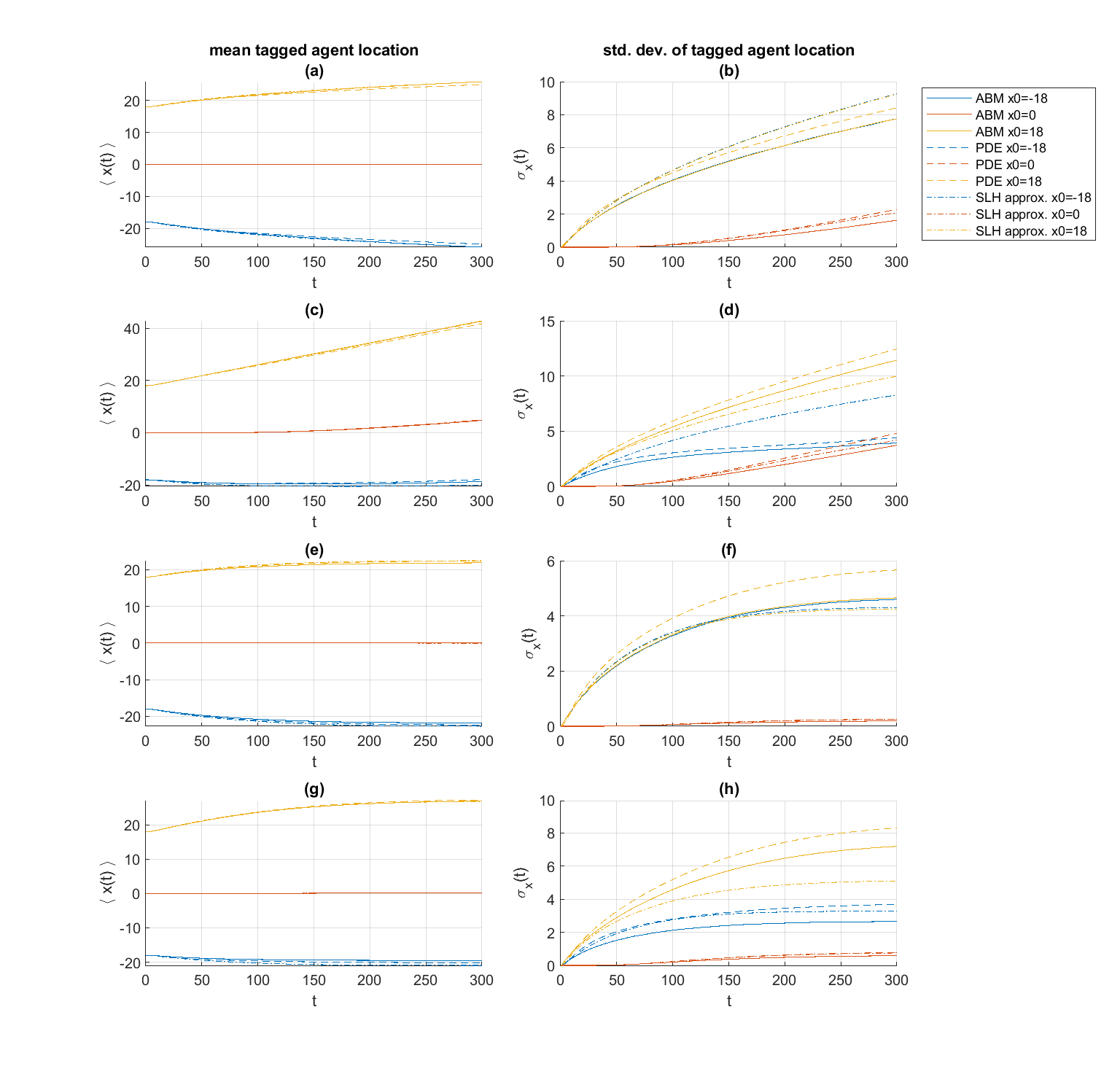}
    \caption{Graphs of the mean $\langle x(t)\rangle$ and standard deviation $\sigma_x(t)$ of tagged agent locations over time in the ABM (solid curves), PDE model (dashed curves) and the SLH approximation~[27] (dot-dash curves), for agents initially located at $x_0=-18$ (blue), $x_0=0$ (red) and $x_0=18$ (yellow). The four rows of plots show the four cases investigated: (a-b) unbiased, no proliferation; (c-d) biased, no proliferation; (e-f) unbiased, with proliferation; (g-h) biased, with proliferation. Note in (b) and (f) the blue and yellow curves for standard deviation coincide almost exactly for all three models due to the symmetry in the model.  }
    \label{fig:S1}
\end{figure}

\newpage 

\begin{figure}
    \centering
    \includegraphics[trim={2.3cm 0 3.2cm 0},clip,width=\linewidth]{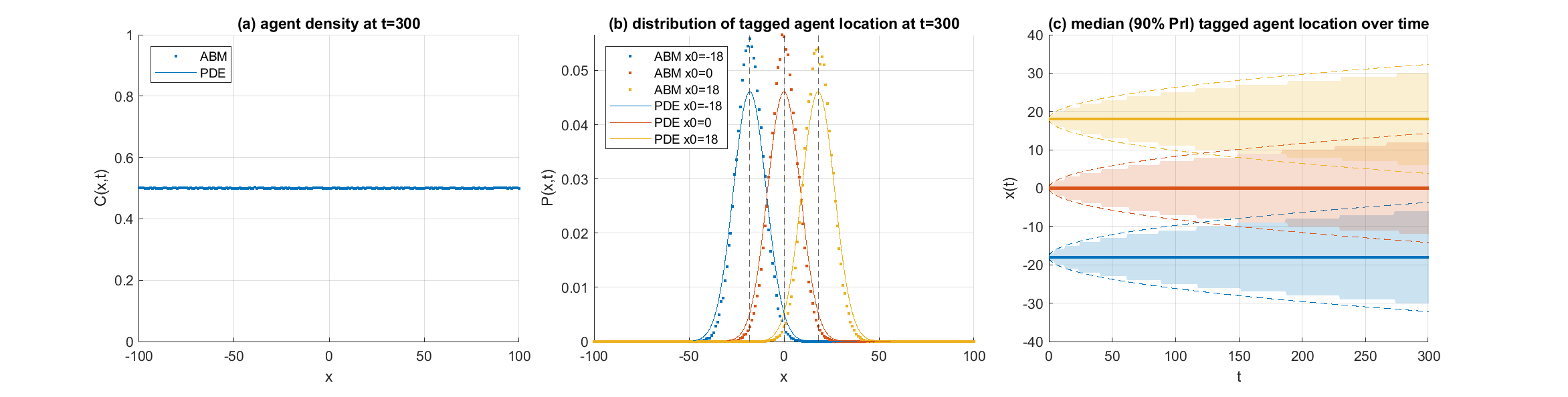}
    \caption{Comparison of ABM and PDE results for a test case in which all lattice sites initially occupied with probability $C_0=0.5$ and with no bias or proliferation. (a) Agent density $C(x,t)$ at $t=300$. (b) Distribution of the location at $t=300$ of tagged agents initially located at $x_0=-18$ (blue), $x_0=0$ (red) and $x_0=18$ (yellow). Vertical dashed lines show the initial location of tagged agents.  (c) Median and 90\% PrI of tagged agent locations as a function of time: ABM results are shown as thick solid curve (median) and shaded band (90\% PrI); PDE results are shown as thin solid curve (median) and dashed curves (90\% PrI). Discrete parameter values $M = 1$ and $\rho_x=Q=0$ corresponding to  $D=0.25$ and $v=\lambda=0$ for a simulation with $\Delta = \tau = 1$. Notice that the PDE solution for $P(x,t)$ slightly overestimates the variance in the distribution of tagged agent locations in the ABM. }
    \label{fig:S2}
\end{figure}

\end{document}